\documentclass[10pt,a4paper,twoside]{article}
\usepackage{epsfig}
\usepackage{baltlat6}
\usepackage{array}
\usepackage{amssymb}
\begin{document}
\ \
\vspace{0.5mm}
\vspace{8mm}

\titlehead{Baltic Astronomy, vol.18, 311--315, 2009}

\titleb{A SEARCH FOR THE INTERMEDIATE SUBGROUP OF THE GAMMA-RAY BURSTS IN THE SWIFT DATASET}

\begin{authorl}
\authorb{D. Huja}{1},
\authorb{J. \v{R}\'{\i}pa}{1}
\end{authorl}

\begin{addressl}
\addressb{1}{Charles University,
Faculty of Mathematics and Physics,
Astronomical Institute,
V Hole\v{s}ovi\v{c}k\'ach 2, 180 00 Prague 8,
Czech Republic;\\
David.HUJA@seznam.cz;\\
ripa@sirrah.troja.mff.cuni.cz}
\end{addressl}

\submitb{Received: 2009 July 20; accepted: 2009 December 1}

\begin{summary}
It has been observed nearly 400 gamma-ray bursts by the Swift satellite.
We search for a third (intermediate) subgroup of the bursts by the standard $\chi^2$ method and F-test.
Supports for the existence of this subgroup are found.
\end{summary}


\resthead{Searching for the intermediate subgroup of the GRBs}
{D. Huja, J. \v{R}\'{\i}pa}

\sectionb{1}{INTRODUCTION}
According to our knowledge, gamma-ray bursts (GRBs) are the most powerful explosions the Universe has ever seen since the Big Bang.
Many papers on the different sky distributions of the different GRB groups were published (e.g., Bal\'azs et al. 1998,
Bal\'azs et al. 1999, M\'esz\'aros et al. 2000a, 2000b, M\'esz\'aros \& \v{S}to\v{c}ek 2003, Vavrek et al. 2008),
on the different phenomena of the short and long GRBs (Bal\'azs et al. 2003, 2004, Fox et al. 2005)
or on the searching for a third (intermediate) GRB subgroup (Horv\'ath 1998, 2002, Mukherjee et al. 1998,
Horv\'ath et al. 2004, 2006, Chattopadhyay et al. 2007, Horv\'ath et al. 2008, Horv\'ath 2009).
With the Swift satellite (Gehrels 2005), since November 20, 2004, we have a tool, which can solve the gamma-ray burst puzzle.
We examine a GRB sample of the Swift catalogue that covers the period November 2004 –- February 2009
(the first/last event is GRB041227/GRB090205) and consists of the 388 GRBs with measured duration.

\sectionb{2}{$\chi^2$ FITTING OF THE GRB DURATIONS}

The first evidence of the existence of three GRB subgroups was found by
Horv\'ath (1998) the by $\chi^2$ fitting (Trumpler et al. 1953; Kendall et
al. 1973) of the duration distribution of the BATSE GRBs.

In Figure 1 there is a distribution of $\log T_{90}$ durations of our data
sample. We created eight histograms of bursts' durations with different
binnings and fitted them by one Gaussian curve (1G), by the sum of two Gaussian
curves (2G), and by the sum of three Gaussian curves (3G).

While fitting with three Gaussian curves, we obtained four fits out of
eight, for which the decrease of $\chi^2$, in comparison to two-Gaussian fit,
is significant ($F\leq5\,\%$). We consider the introduction of a third subgroup
acceptable if the F-test gives the probability $\leq5\,\%$
(Trumpler et al. 1953, Kendall et al. 1973, Band et al. 1997).

In all cases of the different binnings the one Gaussian curve do not fit the
distribution because the goodness-of-fits are $\ll$ 0.01\,\%.
The fits with the sum of two Gaussian curves were in all cases
statistically significant, because their F-tests were less than 5\,\%
(practically less than 0.01\,\%) and the goodness-of-fits were always
higher than 30\,\%.  The introducing of the third subgroup gave the
F-test less than 10\,\%, in some cases less than 5\,\%.
Average best fitted parameters are:
for 2G: $\mu_{1}=0.33\pm0.13$, $\sigma_{1}=0.96\pm0.05$,
$w=0.17\pm0.02$, $\mu_{2}=1.62\pm0.01$, $\sigma_{2}=0.52\pm0.01$; and
for 3G: $\mu_{1}=0.30\pm0.42$, $\sigma_{1}=0.87\pm0.28$,
$w_{1}=0.17\pm0.05$, $\mu_{2}=1.12 \pm0.13$, $\sigma_{2}=0.47\pm0.32$,
$w_{2}=0.32\pm0.12$, $\mu_{3}=1.87\pm0.06$, $\sigma_{3}=0.36\pm0.04$.

\begin{figure}[!tH]
\vbox{
\centerline{\psfig{figure=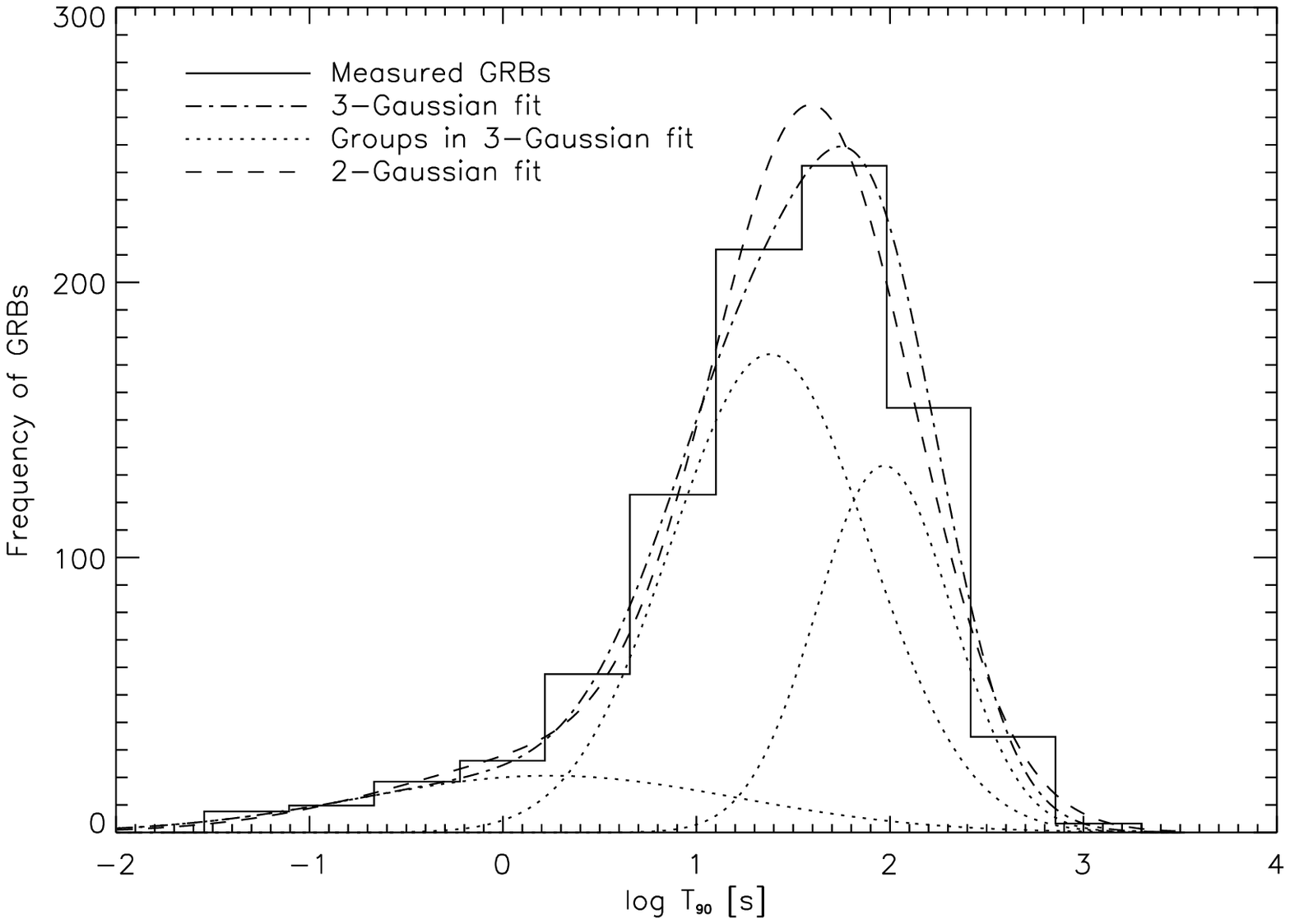,width=91mm,angle=0,clip=}}
\vspace{1mm}
\captionb{1}
{Fitting of the $\log T_{90}$ histogram with
11 bins (fit No.VI.). The number of GRBs per bin is
given by the product of the frequency and the bin width.
The best fits with 2-Gaussian and 3-Gaussian curves are shown.}
}
\end{figure}

\begin{table}[!t]
\begin{center}
\vbox{\footnotesize\tabcolsep=3pt
\parbox[c]{124mm}{\baselineskip=10pt
{\smallbf\ \ Table 1.}{\small\
Results of the $\chi^2$ fitting of the data sample with
388 GRBs.\lstrut}}

\begin{tabular}{lrrrrrrrr}

\hline
Fit         & I.   & II.  & III. &  IV. & V.   & VI.  & VII. & VIII.  \\
No. of      &      &      &      &      &      &      &      &        \\
bins        & 25   & 15   & 31   &   20 & 30   & 11   & 21   & 51     \\
\hline
1 G         &      &      &      &      &      &      &      &        \\
$\chi^2$    & 97.48& 56.59& 94.46& 69.00& 97.51& 65.63& 85.11& 125.34 \\
$\mu$       & 1.45 &  1.47& 1.44 & 1.46 & 1.45 & 1.42 & 1.46 & 1.42   \\
$\sigma$    & 0.93 & 0.83 & 0.89 & 0.87 & 0.90 & 0.86 & 0.88 & 0.90   \\
\hline
2 G         &      &      &      &      &      &      &      &        \\
$\chi^2$    &17.03 & 7.49 & 23.11& 10.86& 27.47& 3.81 & 18.94& 42.00  \\
$\mu_{1}$   & 0.06 & 0.42 & 0.33 &  0.31& 0.25 & 0.47 & 0.38 & 0.43   \\
$\sigma_{1}$& 1.09 & 0.92 & 0.97 &  0.94& 0.95 & 0.94 & 0.95 & 0.95   \\
$\mu_{2}$   & 1.60 & 1.62 & 1.62 &  1.62& 1.61 & 1.61 & 1.62 & 1.63   \\
$\sigma_{2}$& 0.54 & 0.53 & 0.52 &  0.53& 0.52 & 0.51 & 0.52 & 0.51   \\
$w_{2}$     & 0.85 & 0.83 & 0.82 &  0.84& 0.84 & 0.82 & 0.83 & 0.79   \\
$F$\,[\%]   &$10^{-5}$&$10^{-2}$&$10^{-6}$&$10^{-4}$&$10^{-4}$&$10^{-2}$&$10^{-3}$&$10^{-8}$\\
\hline
3 G         &      &      &      &      &      &      &      &        \\
$\chi^2$    &9.99  &  2.38& 16.69& 5.62 & 19.38& 1.72 & 3.73 & 3.27   \\

$\mu_{1}$   &-0.01 & 1.07 &  0.28& -0.28&-0.002& 0.24 & 0.57 & 0.51   \\
$\sigma_{1}$& 1.13 & 0.32 &  0.98&  0.61& 0.83 & 0.98 & 1.09 & 1.01   \\
$w_{1}$     &0.15  & 0.23 & 0.17 &  0.09& 0.13 & 0.13 & 0.20 & 0.22   \\
$\mu_{2}$   & 1.06 &  1.12&  1.06&  0.92& 1.09 & 1.38 & 1.23 & 1.08   \\
$\sigma_{2}$& 0.35 & 1.24 &  0.31& 0.33 & 0.34 & 0.51 & 0.38 & 0.32   \\
$w_{2}$     &0.29  & 0.30 & 0.24 &  0.22&0.31  & 0.57 & 0.39 & 0.23   \\
$\mu_{3}$   & 1.85 &  1.84&  1.84&  1.78& 1.88 & 1.97 & 1.97 & 1.84   \\
$\sigma_{3}$& 0.37 & 0.33 & 0.38 &  0.43& 0.36 & 0.35 & 0.31 & 0.38   \\

$F$[\%]     &{\bf 2.52}&{\bf 3.63}& 5.41&{\bf 4.21}&{\bf 4.94}&5.42&5.90&9.23 \\

\hline
\\
\end{tabular}
}
\end{center}
\vskip-8mm
\end{table}

\sectionb{3}{DISCUSSION}
To discuss the results, first of all, we should remark
that we have proven the existence of the short and long subgroups
in the Swift data-set by the $\chi^2$ method. One Gaussian curve do not
fit the duration. The fit with the sum of three Gaussian ones
is acceptable. It is also highly remarkable that the weight of the short
subgroup is in accordance with the expectation. As it follows from
Horv\'ath et al. (2006), in the BATSE Catalog the populations of the short,
intermediate and long bursts are roughly in the ratio 20:10:70. Nevertheless,
because the short bursts are harder and Swift is more sensitive to
softer GRBs, one may expect that in the Swift database the population
of short GRBs should be  comparable or smaller than 20\,\% due to
instrumental reasons. The obtained weights (being between
10 and 26\,\%) are in accordance with this expectation.
Also the other values of the best fitted parameters - i.e. two
means and two standard deviations - are roughly in the ranges that can be
expected from the BATSE values. The differences can be given by the
different instrumentations. For example, the mean values of the $\log T_{90}$
should be slightly longer in the Swift database compared with the
BATSE data (Barthelmy et al. 2005, Band 2006). In Horv\'ath (1998)
the BATSE means are -0.35 (short) and 1.52 (long), respectively.
Here we obtained for the sample: values from 0.06 to 0.95 (short) and
from 1.60 to 1.63 (long), respectively. All this implies that - concerning the short and long GRBs -
the situation is in essence very similar to the BATSE data-set.

Concerning the third (intermediate) subgroup, our results also supports its existence;
from eight tests four ones gave significances below $5\,\%$.
Hence, strictly speaking, the third subclass does exist and the
probability of the mistake for this claim is not higher than x\,\%,
where 2.52 $<$ x $<$ 9.23. This result is in accordance with the
expectation, once a comparison with the BATSE database is provided.
For the BATSE database the first evidence of third subgroup came from this $\chi^2$ method,
and hence also for the Swift database this test should give positive support for this subgroup,
if the two data-sets are comparable. It is the key result of this article that this expectation is fulfilled.
Our study has shown that the classical $\chi^2$ fitting - in combination with F-test - may
well work also in the Swift database (similarly to the BATSE database (Horv\'ath 1998)).
Horv\'ath et al. (2008) confirmed the third subgroup in the Swift data set by the
maximum likelihood (ML) method. Our significance between 2.52\,\% and
9.23\,\% is higher than the 0.46\,\% significance obtained by Horv\'ath et al.
(2008), which is expectable, because the ML method is a stronger
statistical test. This is seen from two new studies, too: the ML test
on the databases of RHESSI (\v{R}\'{\i}pa et al. 2009) and BeppoSAX (Horv\'ath 2009) satellites, respectively,
confirmed the existence of the third intermediate subclass; on the other hand, the $\chi^2$ test either
did not give a high enough significance for RHESSI data (\v{R}\'{\i}pa et al. 2009) or
was not used for BeppoSAX data at all (Horv\'ath 2009). Comparation of the methods and the results is seen in Table 2.

\begin{table}[!t]
\begin{center}
\vbox{\footnotesize\tabcolsep=3pt
\parbox[c]{124mm}{\baselineskip=10pt
{\smallbf\ \ Table 2.}{\small\
Comparing of the methods and the results by the searching for the third (intermediate) subgroup of the GRBs.
The used methods were: $\chi^2$ method, maximum likelihood method on $T_{90}$ (ML $T_{90}$), maximum likelihood method on
$T_{90}$ vs. hardness ratio (ML $T_{90}$ vs. HR), and the others methods (not specified). As it is seen in this table,
the third (intermediate) subgroup of the GRBs is supported by different methods in each data-set of every satellite
(BATSE, Swift, RHESSI). \lstrut}}

\begin{tabular}{llll}
\\
\hline
Methods         & BATSE                                & Swift                           & RHESSI                               \\
\hline
$\chi^2$        & Yes \tiny(Horv\'ath 1998)            & Yes \tiny(Huja et al. 2009)     & No  ~\tiny(\v{R}\'{\i}pa et al. 2009) \\
ML $T_{90}$     & Yes \tiny(Horv\'ath 2002)            & Yes \tiny(Horv\'ath et al. 2008)& Yes \tiny(\v{R}\'{\i}pa et al. 2009) \\
ML $T_{90}$-HR  & Yes \tiny(Horv\'ath et al. 2006)     & --                              & Yes \tiny(\v{R}\'{\i}pa et al. 2009) \\
Other           & Yes \tiny(Mukherjee et al. 1999,     & --                              & --                                   \\
methods         & ~~~~~~\tiny Chattopadhyay et al. 2007) &                               &                                      \\
\hline
\\
\end{tabular}
}
\end{center}
\vskip-8mm
\end{table}

\sectionb{4}{CONCLUSION}
Since the article Horv\'ath (1998) described the existence of the third (intermediate) subgroup of GRBs in the BATSE database
by the $\chi^2$ fitting of the duration, we worked out an identical procedure on the existing Swift database.
Similarly to the BATSE GRBs, the Swift GRBs do also require (but not as strongly as found by Horv\'ath (1998)) an introduction of the third subgroup.
Our results are very similar to Huja et al. (2009).

The results may be summarized in the following three points:
1. Concerning the short and long subgroups, our results are
in accordance with the expectation: they are also detected in the Swift
database and - in addition - in the Swift database the weight of the
short subgroup is smaller, which can be well explained by the Swift's
higher effective sensitivity to softer bursts.
2. The sample of 388 objects gives support (but not strong enough) for three
subgroups, because from  eight fittings of the whole sample  four ones
confirmed the existence of the intermediate subgroup on a smaller than 5\,\%
significance level. Hence, concerning the Swift database, the
situation is similar to the BATSE data set.
3. Similarly to the BATSE database, here it is shown again that the
classical $\chi^2$ test - in combination with F-test - is also effective for the Swift GRB sample.

\thanks {This study was supported by the GAUK grant No. 46307,
by the Grant Agency of the Czech Republic grant No. 205/08/H005, and
by the Research Program MSM0021620860 of the Ministry of Education
of the Czech Republic. We thank A. M\'esz\'aros for valuable discussion.}

\newpage

\References

\refb Bal\'{a}zs L. G., M\'{e}sz\'{a}ros A., Horv\'{a}th I. 1998, A\&A 339

\refb Bal\'{a}zs L. G., M\'{e}sz\'{a}ros A., Horv\'{a}th I., Vavrek R. 1999, A\&AS, 138, 417

\refb Bal\'{a}zs L. G., Bagoly Z., Horv\'{a}th I. et al. 2003, A\&A, 401, 129

\refb Bal\'{a}zs L. G., Bagoly Z., Horv\'{a}th I. et al. 2004, Baltic Astronomy, 13, 207

\refb Band D. L., Ford L. A., Matteson J. L. et al. 1997, ApJ, 485, 747, Appendix A

\refb Band D. L. 2006, ApJ, 644, 378

\refb Barthelmy S. D., Chincarini G., Burrows D. N. et al. 2005, Nature, 438, 994

\refb Chattopadhyay T., Misra R., Chattopadhyay A. K., Naskar M. 2007, ApJ, 667, 1017

\refb Fox D. B., Frail D. A., Price P. A. et al. 2005, Nature, 437, 845

\refb Gehrels N. et al. 2005,\\
\texttt{http://heasarc.gsfc.nasa.gov/docs/swift/archive/grb\_table}

\refb Horv\'ath I. 1998, ApJ, 508, 757

\refb Horv\'ath I. 2002, A\&A, 392, 791

\refb Horv\'{a}th I., M\'{e}sz\'{a}ros A., Bal\'{a}zs L. G., Bagoly Z.
2004, Baltic Astronomy, 13, 217

\refb Horv\'{a}th I., Bal\'{a}zs L. G., Bagoly Z. et al. 2006, A\&A,
447, 23

\refb Horv\'{a}th I., Bal\'{a}zs L. G., Bagoly Z., Veres P. 2008, A\&A,
489, L1

\refb Horv\'{a}th I. 2009, Ap\&SS, 323, 83

\refb Huja D., M\'{e}sz\'{a}ros A., \v{R}\'{\i}pa J. 2009, A\&A, 504, 67

\refb Kendall M. G., Stuart A. 1973, {\it The Advanced Theory of Statistics},\\ Charles Griffin\&Co. Ltd., London \& High Wycombe

\refb M\'{e}sz\'{a}ros A., Bagoly Z., Vavrek R. 2000a, A\&A, 354, 1

\refb M\'{e}sz\'{a}ros A., Bagoly Z., Horv\'{a}th I. et al. 2000b, ApJ,
539, 98

\refb M\'esz\'aros A., \v{S}to\v{c}ek J. 2003, A\&A, 403, 443

\refb Mukherjee S., Feigelson E. D., Jogesh B. G. et al. 1998, ApJ, 508,
314

\refb \v{R}\'{\i}pa J.,	M\'{e}sz\'{a}ros A., Wigger, C. et al. 2009,
A\&A, 498, 399

\refb Trumpler R. J., Weaver H. F. 1953, {\it Statistical Astronomy},
University of California Press

\refb Vavrek R., Bal\'{a}zs L. G., M\'{e}sz\'{a}ros A. et al. 2008,
MNRAS, 391, 1741

\end{document}